\documentclass[twocolumn]{aastex701}

\pdfoutput=1 
\usepackage{xspace}
\usepackage[encapsulated]{CJK}  
\usepackage{gensymb}
\usepackage{subcaption}
\usepackage{amsmath}

\graphicspath{{Figures/}}

\graphicspath{{./}{Figures/}}

\begin{document}

\title{Characterization of Two Cool Galaxy Outflow Candidates Using Mid-Infrared Emission from Polycyclic Aromatic Hydrocarbons}

\correspondingauthor{Jessica Sutter}
\email{sutterjs@whitman.edu}


\newcommand{\Whitman}{Whitman College, 345 Boyer Avenue, Walla Walla, WA 99362, USA}

\newcommand{\Ox}{Sub-department of Astrophysics, Department of Physics, University of Oxford, Keble Road, Oxford OX1 3RH, UK}

\newcommand{\UGent}{Sterrenkundig Observatorium, Universiteit Gent, Krijgslaan 281 S9, B-9000 Gent, Belgium}

\newcommand{\STScI}{Space Telescope Science Institute, 3700 San Martin Drive, Baltimore, MD 21218, USA}

\newcommand{\MPIA}{Max-Planck-Institut f\"{u}r Astronomie, K\"{o}nigstuhl 17, D-69117, Heidelberg, Germany}

\newcommand{\AURA}{AURA for the European Space Agency (ESA), Space Telescope Science Institute, 3700 San Martin Drive, Baltimore, MD 21218, USA}

\newcommand{\STScIESA}{AURA for the European Space Agency (ESA), Space Telescope Science Institute, 3700 San Martin Drive, Baltimore, MD 21218, USA}

\newcommand{\UCSD}{Department of Astronomy \& Astrophysics, University of California, San Diego, 9500 Gilman Dr., La Jolla, CA 92093, USA}

\newcommand{\JHU}{Department of Physics and Astronomy, The Johns Hopkins University, Baltimore, MD 21218, USA}

\newcommand{\OSU}{Department of Astronomy, The Ohio State University, 140 West 18th Avenue, Columbus, OH 43210, USA}

\newcommand{\CCAPP}{Center for Cosmology and Astroparticle Physics (CCAPP), 191 West Woodruff Avenue, Columbus, OH 43210, USA}

\newcommand{\ARI}{Astronomisches Rechen-Institut, Zentrum f\"{u}r Astronomie der Universit\"{a}t Heidelberg, M\"{o}nchhofstr. 12-14, D-69120 Heidelberg, Germany}

\newcommand{\UCT}{Department of Astronomy, University of Cape Town, Rondebosch 7701, South Africa}

\newcommand{\UConn}{Department of Physics, University of Connecticut, 196A Auditorium Road, Storrs, CT 06269, USA}

\newcommand{\UHawaii}{Institute for Astronomy, University of Hawaii, 2680 Woodlawn Drive, Honolulu, HI 96822, USA}

\newcommand{\UniCA}{Université Côte d'Azur, Observatoire de la Côte d'Azur, CNRS, Laboratoire Lagrange, 06000, Nice, France}

\newcommand{\UAlberta}{Dept. of Physics, University of Alberta, 4-183 CCIS, Edmonton, Alberta, T6G 2E1, Canada}

\newcommand{\Arcetri}{INAF — Osservatorio Astrofisico di Arcetri, Largo E. Fermi 5, I-50125, Florence, Italy}

\newcommand{\UWyoming}{Department of Physics and Astronomy, University of Wyoming, Laramie, WY 82071, USA}

\newcommand{\LJMU}{Astrophysics Research Institute, Liverpool John Moores University, 146 Brownlow Hill, Liverpool L3 5RF, UK}

\newcommand{\ITA}{Universit\"{a}t Heidelberg, Zentrum f\"{u}r Astronomie, Institut f\"{u}r Theoretische Astrophysik, Albert-Ueberle-Str 2, D-69120 Heidelberg, Germany}

\newcommand{\CfA}{Center for Astrophysics $\mid$ Harvard \& Smithsonian, 60 Garden St., 02138 Cambridge, MA, USA}

\newcommand{\MPE}{Max-Planck-Institut f\"{u}r Extraterrestrische Physik (MPE), Giessenbachstr. 1, D-85748 Garching, Germany}

\newcommand{\Surrey}{Department of Physics, University of Surrey, Guildford GU2 7XH, UK}

\newcommand{\ESO}{European Southern Observatory, Karl-Schwarzschild Stra{\ss}e 2, D-85748 Garching bei M\"{u}nchen, Germany}

\newcommand{\IWR}{Universit\"{a}t Heidelberg, Interdisziplin\"{a}res Zentrum f\"{u}r Wissenschaftliches Rechnen, Im Neuenheimer Feld 205, D-69120 Heidelberg, Germany}

\newcommand{\ulyon}{Univ Lyon, Univ Lyon1, ENS de Lyon, CNRS, Centre de Recherche Astrophysique de Lyon UMR5574, F-69230 Saint-Genis-Laval France}

\newcommand{\COOL}{Cosmic Origins Of Life (COOL) Research DAO, coolresearch.io}

\newcommand{\OAN}{Observatorio Astron{\'o}mico Nacional (IGN), C/ Alfonso XII 3, E-28014 Madrid, Spain}

\newcommand{\UBonn}{Argelander-Institut f\"{u}r Astronomie, Universit\"{a}t Bonn, Auf dem H\"{u}gel 71, 53121 Bonn, Germany}

\newcommand{\kipac}{Kavli Institute for Particle Astrophysics \& Cosmology (KIPAC), Stanford University, CA 94305, USA}

\newcommand{\Umanc}{Jodrell Bank Centre for Astrophysics, Department of Physics and Astronomy, University of Manchester, Oxford Road, Manchester M13 9PL, UK}

\newcommand{\NRAO}{National Radio Astronomy Observatory, 520 Edgemont Road, Charlottesville, VA 22903, USA}

\newcommand{\ANU}{Research School of Astronomy and Astrophysics, Australian National University, Canberra, ACT 2611, Australia}

\newcommand{\AThreeD}{ARC Centre of Excellence for All Sky Astrophysics in 3 Dimensions (ASTRO 3D), Australia}

\newcommand{\HD}{\label{HD} Astronomisches Rechen-Institut, Zentrum f\"{u}r Astronomie der Universit\"{a}t Heidelberg, M\"{o}nchhofstra\ss e 12-14, D-69120 Heidelberg, Germany}

\newcommand{\IAC}{Instituto de Astrof\'isica de Canarias, C/ V\'ia L\'actea s/n, E-38205, La Laguna, Spain}

\newcommand{\ULL}{Departamento de Astrof\'isica, Universidad de La Laguna, Av. del Astrof\'isico Francisco S\'anchez s/n, E-38206, La Laguna, Spain}

\newcommand{\Princeton}{Department of Astrophysical Sciences, Princeton University, 4 Ivy Lane, Princeton, NJ 08544, USA}

\newcommand{\IRAM}{IRAM, 300 rue de la Piscine, 38400 Saint Martin d'H\'{e}res, France}

\newcommand{\LERMA}{LERMA, Observatoire de Paris, PSL Research University, CNRS, Sorbonne Universit\'{e}s, 75014 Paris}

\newcommand{\YB}{Centro de Desarrollos Tecnol\'ogicos, Observatorio de Yebes (IGN), 19141 Yebes, Guadalajara, Spain}

\newcommand{\uwa}{International Centre for Radio Astronomy Research, University of Western Australia, 7 Fairway, Crawley, 6009, WA, Australia}

\newcommand{\UMD}{Department of Astronomy and Joint Space-Science Institute, University of Maryland, College Park, MD 20742, USA}

\newcommand{\NOIRLab}{International Gemini Observatory/NSF NOIRLab, 950 N. Cherry Avenue, Tucson, AZ 85719, USA}

\author[0000-0002-9183-8102]{Jessica Sutter}
\email[]{sutterjs@whitman.edu}
\affiliation{\Whitman}

\author[0000-0002-4378-8534]{Karin Sandstrom}
\affiliation{\UCSD}
\email[]{kmsandstrom@ucsd.edu}

\author[0000-0001-8241-7704]{Ryan Chown}
\affiliation{\OSU}
\email[]{chown.5@osu.edu}

\author[0000-0002-4755-118X]{Oleg Egorov}
\affiliation{\HD}
\email[]{oleg.egorov@uni-heidelberg.de}

\author[0000-0002-2545-1700]{Adam~K.~Leroy}
\affiliation{\OSU}
\affiliation{\CCAPP}
\email[]{leroy.42@osu.edu}

\author[0000-0002-5235-5589]{J\'{e}r\'{e}my Chastenet}
\affiliation{\UGent}
\email[]{jeremy.chastenet@ugent.be}

\author[0000-0002-5480-5686]{Alberto D. Bolatto}
\affiliation{\UMD}
\email[]{bolatto@umd.edu}

\author[0000-0002-0012-2142]{Thomas~G.~Williams}
\affiliation{\Ox}
\email[]{thomas.williams@physics.ox.ac.uk}

\author[0000-0002-5782-9093]{Daniel~A.~Dale}
\affiliation{Department of Physics and Astronomy, University of Wyoming, Laramie, WY 82071, USA}
\email[]{ddale@uwyo.edu}

\author[0000-0002-8553-1964]{Amirnezam Amiri}
\affiliation{Department of Physics, University of Arkansas, 226 Physics Building, 825 West Dickson Street, Fayetteville, AR 72701, USA}
\email[]{amirnezamamiri@gmail.com}

\author[0000-0003-0946-6176]{Médéric Boquien}
\affiliation{\UniCA}
\email[]{mederic.boquien@oca.eu}

\author[0000-0001-5301-1326]{Yixian Cao}
\affiliation{Max-Planck-Institut f\"ur Extraterrestrische Physik (MPE), Giessenbachstr. 1, D-85748 Garching, Germany}
\email[]{ycao@mpe.mpg.de}

\author[0000-0002-2885-6172]{Simthembile Dlamini}
\affiliation{\UCT}
\email[]{simther4111@gmail.com}

\author[0000-0002-6155-7166]{\'{E}ric Emsellem}
\affiliation{\ESO}
\affiliation{\ulyon}
\email[]{eric.emsellem@eso.org}

\author[0000-0002-1370-6964]{Hsi-An Pan}
\affiliation{Department of Physics, Tamkang University, No.151, Yingzhuan Road, Tamsui District, New Taipei City 251301, Taiwan} 
\email[]{hapan@gms.tku.edu.tw}

\author[0000-0003-2721-487X]{Debosmita Pathak}
\affiliation{Department of Astronomy, Ohio State University, 180 W. 18th Ave, Columbus, Ohio 43210}
\affiliation{Center for Cosmology and Astroparticle Physics, 191 West Woodruff Avenue, Columbus, OH 43210, USA}
\email[]{pathak.89@buckeyemail.osu.edu}

\author[0000-0003-4770-688X]{Hwihyun~Kim}
\affiliation{\NOIRLab}
\email[]{hwihyun.kim@noirlab.edu}

\author[0000-0002-0560-3172]{Ralf S.\ Klessen}
\affiliation{Universit\"{a}t Heidelberg, Zentrum f\"{u}r Astronomie, Institut f\"{u}r Theoretische Astrophysik, Albert-Ueberle-Str.\ 2, 69120 Heidelberg, Germany}
\affiliation{Universit\"{a}t Heidelberg, Interdisziplin\"{a}res Zentrum f\"{u}r Wissenschaftliches Rechnen, Im Neuenheimer Feld 225, 69120 Heidelberg, Germany}
\affiliation{Harvard-Smithsonian Center for Astrophysics, 60 Garden Street, Cambridge, MA 02138, USA}
\affiliation{Elizabeth S. and Richard M. Cashin Fellow at the Radcliffe Institute for Advanced Studies at Harvard University, 10 Garden Street, Cambridge, MA 02138, USA}
\email[]{klessen@uni-heidelberg.de}

\author[0009-0001-5949-1524]{Hannah~Koziol}
\affiliation{\UCSD}
\email[]{hkoziol@ucsd.edu}

\author[0000-0002-5204-2259]{Erik Rosolowsky}
\affiliation{\UAlberta}
\email[]{rosolowsky@ualberta.ca}

\author[0000-0002-6313-4597]{Sumit K. Sarbadhicary}
\affiliation{Department of Physics and Astronomy, The Johns Hopkins University, Baltimore, MD 21218 USA}
\email[]{ssarbad1@jh.edu}

\author[0000-0002-3933-7677]{Eva~Schinnerer}
\affiliation{\MPIA}
\email[]{schinner@mpia.de}

\author[0000-0002-8528-7340]{David~A.~Thilker}
\affiliation{\JHU}
\email[]{dthilker@pha.jhu.edu}

\author[0000-0001-7130-2880]{Leonardo \'Ubeda}
\affiliation{Space Telescope Science Insititue, Baltimore, Maryland 21218}
\email[]{lubeda@stsci.edu}

\author[0009-0005-8923-558X]{Tony Weinbeck}
\affiliation{Department of Physics and Astronomy, University of Wyoming, Laramie, WY 82071, USA}
\email[]{}




\begin{abstract}
We characterize two candidate cool galactic outflows in two relatively low mass, highly inclined Virgo cluster galaxies: NGC\,4424 and NGC\,4694. Previous analyses of observations using the Atacama Large Millimeter Array (ALMA) carbon monoxide (CO) line emission maps did not classify these sources as cool outflow hosts.  Using new high sensitivity, high spatial resolution, JWST mid-infrared photometry in the polycyclic aromatic hydrocarbon (PAH)-tracing F770W band, we identify extended structures present off of the stellar disk. The identified structures are bright in the MIRI F770W and F2100W bands, suggesting they include PAHs as well as other dust grains. As PAHs have been shown to be destroyed in hot, ionized gas, these structures are likely to be outflows of cool ($T \leq 10^4$~K) gas. This work represents an exciting possibility for using mid infrared observations to identify and measure outflows in lower mass, lower star formation galaxies.

\end{abstract}



\section{Introduction} \label{sec:intro}

\begin{figure*}
\centering
\includegraphics[width=\linewidth]{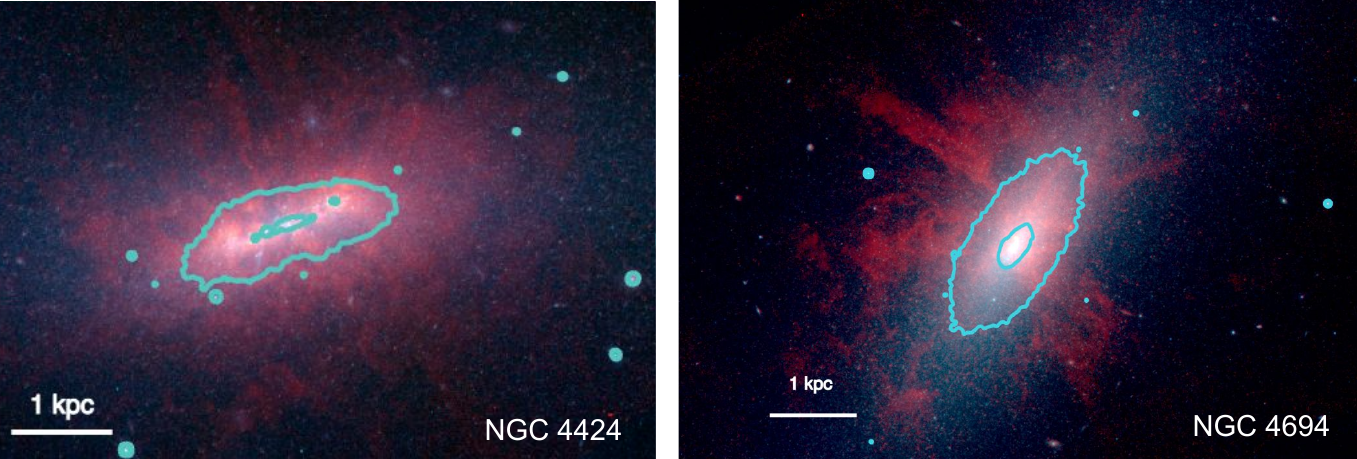} 
\caption{Three-color images of the two outflow candidate galaxies, NGC\,4424 (left) and NGC\,4694 (right).  In both, red is F770W (PAH emission), green is F335M (stellar + PAH emission), and blue is F300M (stellar emission).  Cyan contours represent the 10~MJy/sr and 2~MJy/sr in F300M, highlighting the location and extent of the stellar disk.  Gray scale bars in the lower left of each image show 1~kpc.  North is up, and east is to the right in both images.}
\label{fig:rgb}
\end{figure*}

Outflows --- gas and dust ejected from the disk of a galaxy --- are an important factor in driving galaxy evolution and quenching.  High star formation activity or an active galactic nucleus (AGN) can generate winds powerful enough to create outflows \citep{Heckman2002, Fabian2012}, contributing to the eventual suppression of star formation \citep[see e.g.,][]{Veilleux2020, Thompson2024}.  Outflows also enrich the gas in the circumgalactic and intracluster media (CGM and ICM).  The exact composition, structure, and mass of outflowing material can provide important information about how galaxies and the surrounding medium evolve over time. Millimeter-wavelength observations have shown that these outflows can contain cool, molecular gas \citep{Cicone2014}. As cool gas \citep[$T \lesssim 10^4$~K, i.e. the temperatures at which hydrogen will be neutral, and therefore PAHs are most likely to be found;][]{Sandstrom2012, Hensley2022, Chown2024} plays an essential role in forming new stars, its removal and recycling has major implications for galaxy evolution.

Cool outflows can carry small dust grains, including polycyclic aromatic hydrocarbons (PAHs), out of the host galaxy \citep{Engelbracht2006, Irwin2006, Irwin2007, McCormick2013}.  PAHs produce many near and mid-infrared emission features, which can be responsible for up to 20\% of the infrared emission from a galaxy \citep{Smith2007}, and have been shown to trace dense gas \citep{Chown2025}.  The near and mid-infrared observations required to trace the small dust grains in these past studies were primarily made by the \textit{Spitzer Space Telescope}, with additional observations from the \textit{Infrared Space Observatory} (ISO) and \textit{Akari}.  JWST provides an unprecedented opportunity to trace the PAHs and other dust grains in outflows and generally in extraplanar material at unprecedented sensitivity and spatial resolution \citep[see, e.g.,][]{Bolatto2024, Chastenet2024, Fisher2024}, improving on both the sensitivity and spatial resolution of \textit{Spitzer} by a factor of $\sim10$.  

We identify and characterize two candidate galactic-scale cool outflows detected in the mid-infrared (MIR) with JWST in galaxies NGC\,4424 and NGC\,4694.  While finding PAHs in outflowing gas is not unusual \citep[see, e.g.,][which finds extended PAH emission in 15/16 galaxies with known outflows]{McCormick2013}, it is rare that outflowing gas is primarily identified or characterized using PAH emission.  Both galaxies are relatively low mass \citep[log$_{10}M_\star/M_\odot \sim 9.9$,][]{Leroy2019}  Virgo Cluster occupants. While both galaxies were included in the work of \citet{Stuber2021}, which used velocity profiles from ALMA CO 2-1 maps of the 90 PHANGS-ALMA galaxies to identify molecular outflows, neither were confirmed as an outflow host (see Section~\ref{sec:mw_comp}). The presence of outflowing gas is evident in the filamentary PAH emission detected in the MIR, which extends well beyond the stellar disk observed in the near-infrared.  As we only use photometry and no velocity measurements for this characterization, we cannot definitively confirm whether these structures are truly outflows.  Because of this, we refer to the detected structures as `candidate' outflows throughout the paper.  Three-color images highlighting the differences in the stellar and dust disks of both NGC\,4424 and NGC\,4694 are displayed in Figure~\ref{fig:rgb}.  Properties of both galaxies are listed in Table~\ref{tab:galprops}

\section{Outflow Candidates and Data}
\label{sec:data}

\subsection{NGC\,4424}
\begin{figure*}[t]
    \centering
    \includegraphics[width=\linewidth]{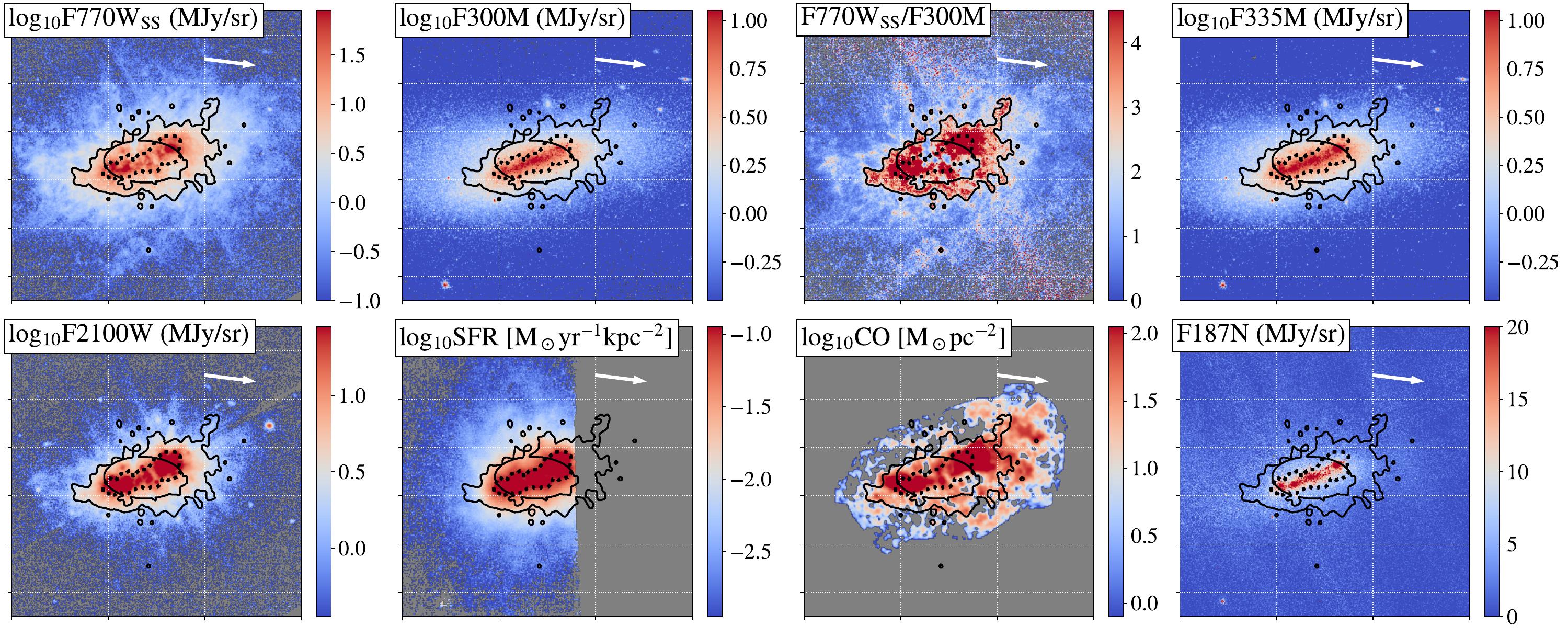}
    \caption{Multi-wavelength images of NGC\,4424. In each panel, the white arrow represents the direction toward M87, and the dashed black contour outlines areas where the star formation rate surface density exceeds 0.1~M$_\odot$~yr$^{-1}$~kpc$^{-2}$; i.e., the Heckman Number \citep{Heckman2002}.  The HI tail extends in the opposite direction as M87, opposite the white arrow, but is outside the field of view of these images so is not shown here. Solid black contours represent areas identified as outflows and used to estimate the outflow rates in Section~\ref{sec:out_rate}.  In all panels, north is up and east is to the right. \textit{Top Row, Left to right:} F770W$_{\rm{SS}}$, highlighting mid-infrared PAH emission, F300M, showing the stellar disk, F770W$_{\rm{PAH}}$/F300M, emphasizing the differences between the stellar disk and the extended PAH emission, F335M, showing the lack of enhanced emission from small PAHs. \textit{Bottom Row, Left to right:} F2100W, showing the emission from larger dust grains, SFR measured by extinction corrected H$\alpha$ from MUSE, and CO (2-1) integrated intensity from ALMA, and F187N map, showing the lack of Paschen~$\alpha$ emission in outflows.  The MUSE coverage is incomplete, so only the area of the galaxy with coverage is shown.}
    \label{fig:MW_ngc4424}
\end{figure*}

NGC\,4424, shown in Figure~\ref{fig:MW_ngc4424}, is classified as a peculiar starburst galaxy due to the heart-shaped isophotes measured in the R band by \citet{Cortes2006}. It is located near the edge of the Virgo cluster (projected distance of 3.11\textdegree or 880~kpc from M87, see $\theta_{\rm M87}$ in Table~\ref{tab:galprops}), but has a projected distance of $\sim 1.5\degree$ (420~kpc) from M49, another giant elliptical in Virgo.  This suggests that NGC\,4424 may be part of a smaller cluster currently in the process of merging into the larger Virgo cluster \citep{Mei2007}.  Optical studies of NGC\,4424 have suggested the possibility of a past intermediate-mass merger based on asymmetries in the stellar disk \citep{Cortes2006}, along with an ionized gas outflow noted in the asymmetries of the [\ion{O}{3}] velocity as well as an extended H$\alpha$ tail noted in \citet{Boselli2018}.  Additional radio continuum measurements show two polarized cones, indicative of a nuclear outflow \citep{Vollmer2013}.  Both these works suggest the origin of these outflows was a central starburst approximately 500~Myr ago, triggered by the past collisions.  They support this claim through measurements of the velocity dispersion in the ionized gas outflows ratios of emission lines suggestive of a starburst \citep{Boselli2018}.  This past starburst activity is likely the source of the filamentary PAH emission structures presented here. \ion{H}{1} data show the presence of a tidally stripped tail \citep{Lee2022}.

\subsection{NGC\,4694}
NGC\,4694, shown in Figure~\ref{fig:MW_ngc4694}, is a lenticular galaxy similarly located at the edge of the Virgo cluster, far from M87 (projected distance of  $4.58\degree$ or 1.3~Mpc).  While there is no clear evidence of past mergers, \ion{H}{1} observations have detected the presence of a gas bridge connecting it to VCC2062, a nearby dwarf galaxy \citep{Lisenfeld2016}, while \citet{Lee2022} classify it as currently undergoing tidal stripping based on the shape and distribution of the \ion{H}{1} tail.  The \citet{VeronCetty2010} catalog identifies NGC\,4694 as the host of an active galactic nucleus, which could be the source of the presented outflow candidate.

\subsection{James Webb Space Telescope Imaging}
The two outflow candidates were identified based on unusual structures discerned from mid-infrared observations obtained by JWST.  Both of these galaxies are part of the Physics at High Angular resolution in Nearby GalaxieS (PHANGS)-JWST Cycle 2 Treasury \citep[GO 3707, PI: A. Leroy, see appendix in][]{Chown2025}, which mapped them in two MIRI and four NIRCam bands.  The two MIRI bands are F770W (central wavelength 7.7~$\mu$m, width 1.95~$\mu$m, resolution 0\farcs27), which includes the 7.7~$\mu$m PAH feature and is therefore used to trace PAH emission, and F2100W  (central wavelength 21.0~$\mu$m, width 4.58~$\mu$m, resolution 0\farcs67), which is dominated by stochastically-heated small dust grains.  The four NIRCam bands are F150W (central wavelength 1.5~$\mu$m, width 0.318~$\mu$m, resolution 0\farcs049) primarily tracing stellar continuum, F187N (1.87~$\mu$m, 0.024~$\mu$m width resolution 0\farcs61), covering the Paschen~$\alpha$ emission line, F300M (3.0~$\mu$m, 0.318~$\mu$m width, resolution 0\farcs097), primarily tracing warm, small dust grain and stellar continuum emission, and F335M (central wavelength 3.35~$\mu$m, 0.348$\mu$m width angular resolution 0\farcs109), which includes the 3.3~$\mu$m PAH feature.  For this work, we focus primarily on the MIRI bands, where the observed outflow is most prominent. We use the F300M band for comparison as it shows the location of the stellar disk.  All JWST data were processed using \texttt{pjpipe}\footnote{https://pjpipe.readthedocs.io/en/latest/} \citep{Williams2024}, which supplements the official JWST pipeline.  \texttt{pjpipe} modifies the official pipeline to suit observations of large extended sources through additional adjustments to the astrometric alignment and the background matching.  As these images often have few point sources identified in the mid-infrared, an additional alignment step is added to correct for small offsets that remain when using the standard pipeline.  Additionally, as these images typically have large scale diffuse emission from the source that fills the field, \texttt{pjpipe} replaces the \texttt{skymatch} algorithm used for background matching between MIRI tiles in the standard pipeline with a pixel-by-pixel method that takes into account the wide-spread signal that exists in the overlap regions.  Finally, \texttt{pjpipe} adds an additional absolute flux calibration step called ``anchoring'' in which the images are matched to wider field of view archival images taken at similar wavelength from either IRAC or WISE to determine the true background values \citep[see ][for more details]{Williams2024}.  All MIRI observations have dedicated background images observed in parallel to the NIRCam observations, which are used to remove background signals during level two of the pipeline. To match the resolution between the different filters, we smooth all data to a common resolution of 0\farcs85, which corresponds to a physical resolution of $\sim65$~pc in both candidates.  This has been shown to improve the signal to noise ratio in low flux regions without a major loss to the high spatial sensitivity provided by JWST \citep{Williams2024}.  The data were smoothed using the \texttt{stpsf} models and the method described in \citet{Aniano2011}.  With this smoothing method, the assumed 1$\sigma$ sensitivities in each band are: 0.13 MJy~sr$^{-1}$ for F770W, 0.27 MJy~sr$^{-1}$ for F2100W, 0.1 MJy~sr$^{-1}$ for F150W, 1 MJy~sr$^{-1}$ for F187N, 0.052 MJy~sr$^{-1}$ for F300M, and 0.050 MJy~sr$^{-1}$ for F335M.

While the F770W filter is dominated by PAH and dust emission, in many sight lines the Rayleigh-Jeans tail of stellar continuum emission contributes a significant fraction of the observed F770W intensity.  To obtain the F770W intensity solely from dust and PAH emission, we subtract stellar continuum using the calibration from  \citet{Sutter2024}:
$$\rm{F770W}_{\rm{SS}}=\rm{F770W}-0.22\times \rm{F300M}.$$
We do not remove any other underlying continuum, which would likely be from larger dust grains, as both PAHs and larger dust grains imply cooler gas in the outflows so emission from these grains is of interest in our analysis.  We use the F300M images to remove continuum instead of F150W due to persistent striping in F150W which we have yet to model and remove, and can dominate the emission in this filter in low signal to noise regions.

\subsection{Multi-Wavelength Data}
We use data from several archival datasets of these two sources.  Specifically, we have optical integral field unit (IFU) data from the Multi-Unit Spectral Explorer (MUSE) on the Very Large Telescope (VLT) obtained as part of programs 097.D-0408, PI J. Anderson (NGC~4424) and 110.244E, PI L. Cortese (NGC~4694), Atacama Large Millimeter Array (ALMA) maps of the CO ({\it J}=2-1) line obtained as part of PHANGS program \citep{Leroy2021a, Leroy2021b} , and 21~cm \ion{H}{1} maps from the Very Large Array (VLA) Imaging of Virgo spirals in Atomic gas (VIVA) survey \citep{Chung2009}. Together, these data provide a view of the stars and ionized gas (MUSE), the molecular gas (ALMA), and the neutral gas (VIVA). The MUSE data are used to measure the star-formation rate surface density, using the extinction-corrected H$\alpha$ data and the method described in \citet{Belfiore2023} and are reduced using the method described in \citet{Emsellem2022}.

\begin{deluxetable}{lcc}
\tablecolumns{3}
    \caption{Properties of Host Galaxies and the Outflow Candidates}
    \tablehead{ \colhead{Property} & \colhead{NGC\,4424} & \colhead{NGC\,4694} }
    \startdata
    Distance [Mpc] & 16.20 & 15.76 \\
    Inclination [deg] & 58.2 & 60.7 \\
    Stellar Mass [M$_\odot$] & $8.1\times10^9$ & $7.2\times10^9$\\
    Position Angle [deg] & 88 & 143 \\
    $R_{25}$ [arcmin] & 1.5 & 1.0 \\
    log$_{10}$M$_\star$ [M$_\odot$] & 9.9 & 9.9 \\
    SFR$_{\rm{z0mgs}}$\tablenotemark{a} [M$_\odot$ yr$^{-1}$]& 0.30 & 0.15 \\
    SFR$_{\rm{MUSE}}$ [M$_\odot$ yr$^{-1}$] & 0.18 & 0.06 \\
    $\theta_{\rm{M87}}$\tablenotemark{b} [deg] & 3.11 & 4.58 \\
    \cutinhead{Properties of Outflows} 
    Extent [kpc] & 0.73--1.1 & 1.2--2.3 \\
    Estimated Gas Mass [$\log_{10}$M$_\odot$] & $7.11^{7.85}_{6.76}$ & $6.37^{7.10}_{6.01}$ \\
    $\overline{\rm{F770W_{\rm{SS}} / F2100W}}$ & 1.45 & 1.38 \\
    Estimated $\dot{M}$ [M$_\odot$ yr$^{-1}$] & $0.4^{+2.0}_{-0.3}$ & $0.2^{+0.8}_{-0.15}$ \\
    \enddata
      \tablecomments{Distances are from \citet{Anand2021}, orientation parameters are from \citet{Lang2020} which used rotation curves to measure inclinations, PAs, and central RA and Declinations, and the stellar masses and star formation rates are from \citet{Leroy2019}. Outflow candidate properties were determined using the method described in Section~\ref{sec:discuss} and the contours shown in Figures~\ref{fig:MW_ngc4424} and \ref{fig:MW_ngc4694}. \label{tab:galprops}} 
      \tablenotetext{a}{Star formation rates from the z=0 Multi-wavelength Galaxy Synthesis are global measurements, covering a larger area than the MUSE maps, which is why they are higher than the SFR values from MUSE.}
      \tablenotetext{b}{Angular separation from M87, i.e. the center of the Virgo Cluster, from \citet{Lee2022}.}
\end{deluxetable}

\begin{figure*}
    \centering
    \includegraphics[width=\linewidth]{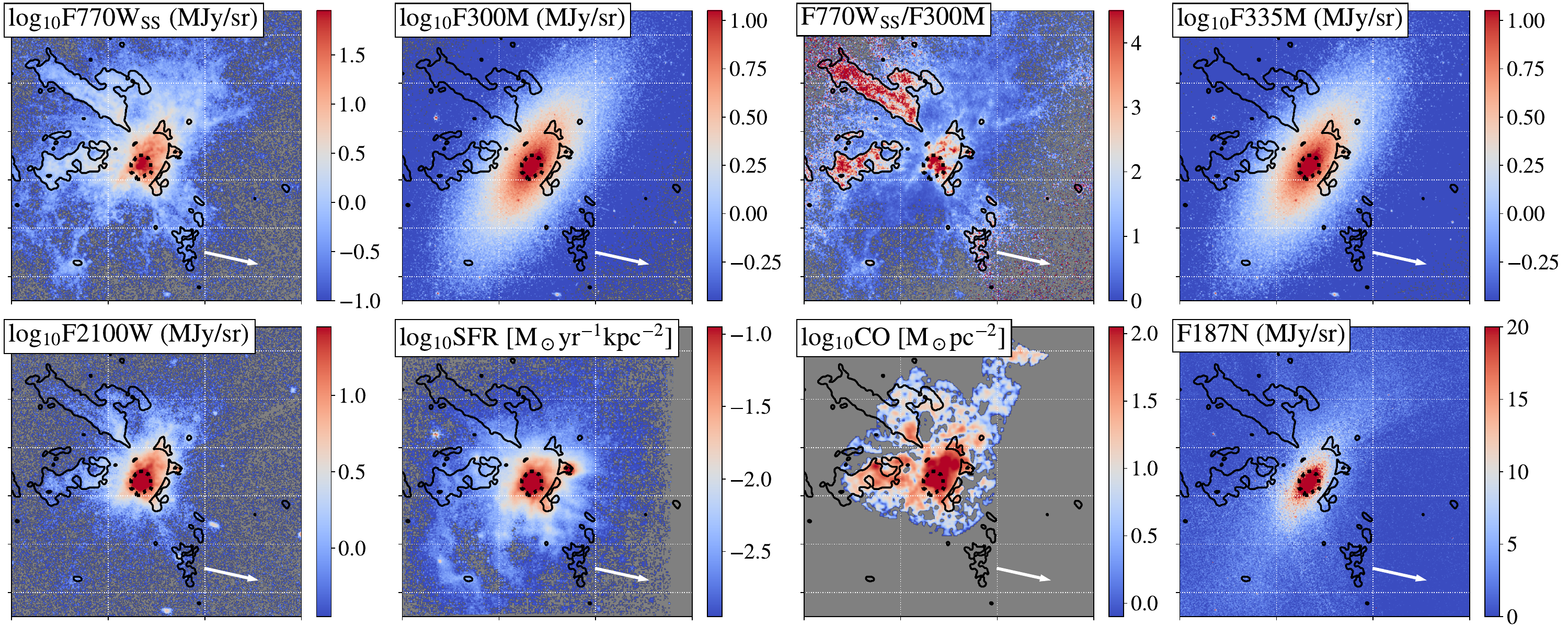}
    \caption{Same as Figure~\ref{fig:MW_ngc4424}, but for NGC\,4694.  Similarly, emission in the mid-infrared tracing dust is extended off the stellar plane. Similar to NGC\,4424, the HI tail extends in the opposite direction as the arrow pointing towards M87, but is outside of the field of view of these images.}
    \label{fig:MW_ngc4694}
\end{figure*}

\section{Outflow Candidate Identification}
\label{sec:id}
The outflow candidates were both identified as part of routine quality assurance (QA) checks of all the PHANGS-JWST Cycle 2 Treasury sources. As part of the QA process, NIRCam and MIRI images are compared and overlaid to check for astrometry shifts across the images.  During the completion of these checks for all 74 galaxies in the survey, NGC\,4424 and NGC\,4694 alone were identified as having clearly distinct structures in the near-infrared bands, tracing primarily starlight and the mid-infrared bands, tracing primarily dust emission.  This check includes a step where ratio maps of the NIRCam and MIRI data are produced and examined, primarily for checking for astrometry shifts. The ratio maps showed elevated levels of dust emission perpendicular to the stellar disk, indicating the presence of an outflow.  Only these two in the sample of 74 galaxies show these structures.

Based on these notes, further inspection of these two sources was performed, identifying the filamentary structures extending off of both sides of the disk.  To isolate the pixels that are part of the filamentary structures we use the ratio of the F770W$_{\rm{SS}}$ and F300M filters.  This ratio is chosen as F770W$_{\rm{SS}}$ is our only tracer of PAH emission and F300M traces the stellar disk.  We use F300M instead of the F150W NIRCam images due to prominent striping patterns in the F150W data that become the dominant signal at high radius and cause anomalously low values of the F770W$_{\rm{SS}}$/F150W value. Contours encapsulating the filaments were produced by identifying all pixels with a F770W$_{\rm{SS}}$/F300M value greater than twice the average F770W$_{\rm{SS}}$/F300M value all pixels within the central 0.15$R_{25}$, and a projected radius of greater than 0.15$R_{25}$ to exclude the region used to find the central value.  The value of 2$\times$ the central 0.15$R_{25}$ $\overline{\rm{F770W}_{\rm{SS}}/\rm{F300M}}$ was determined using comparisons to the 18 additional galaxies included in the PHANGS-JWST Treasuries with $i>55$\textdegree, shown in the upper panels of Figure~\ref{fig:Comparison}.  The identified structures are outlined with black contours in Figures~\ref{fig:MW_ngc4424} and \ref{fig:MW_ngc4694}.  Further details about the comparison sample are provided in Appendix~\ref{app:comp_sample}.  Only the two outflow candidate host galaxies show F770W$_{\rm{SS}}$/F300M reaching above twice the central ratio, indicating that this value is a good indicator of the unique conditions in cool outflows. 

\subsection{Distinguishing Between Outflows and Tidal Stripping}
\label{sec:discuss}

Both NGC\,4424 and NGC\,4694 have been identified as galaxies experiencing ram pressure stripping based on 21 cm observations of 52 Virgo Cluster galaxies \citep{Lee2022}.  Interestingly, NGC\,4424 and NGC\,4694 were the only two galaxies in this sample with \ion{H}{1} tails classified as `atypical,' where the predicted ram pressure would not be enough to produce the large extent of the observed tidal tail \citep{Lee2022}.  With this finding in mind, we carefully consider whether our observations are indeed indicative of an outflow or are instead the PAH component of these stripped tails.  The primary factor that indicates that we are observing outflowing gas, as opposed to stripped gas, is the presence of the structures above and below the stellar disk in both candidate galaxies, which would likely not be the case if gas were being stripped.  These structures are approximately an order of magnitude smaller in extent than the detected stripped gas tails and do not appear to be associated with them (see Section~\ref{sec:mw_comp} for further discussion).

\begin{figure*}
    \centering
    \includegraphics[width=\linewidth]{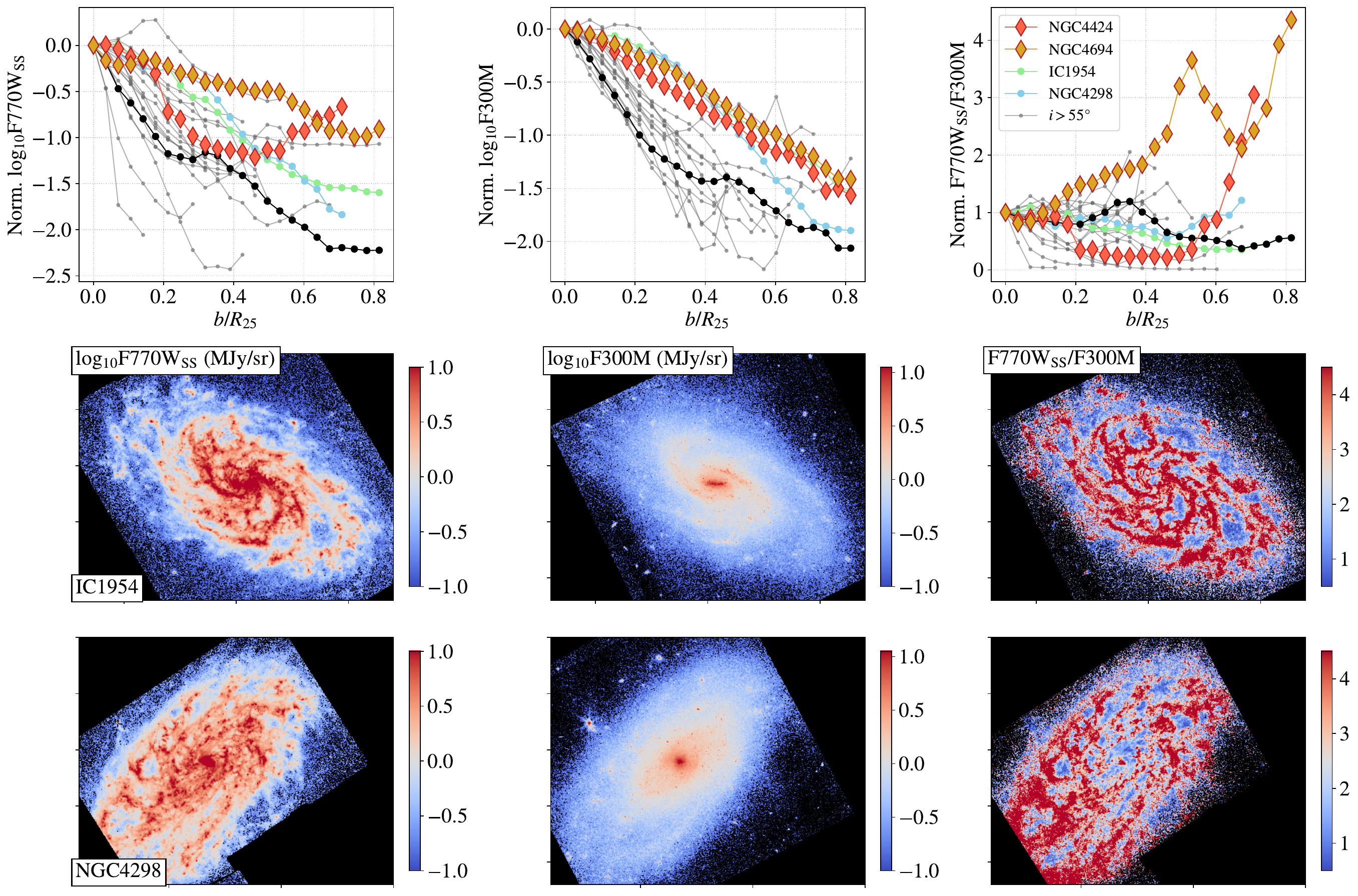}
    \caption{\textit{Top Row:} Profiles of the median flux normalized to the central value, shown in the direction along the minor axis across the full image.  The left panel shows the F770W$_{\rm SS}$ data, the middle panel shows F300M tracing the stars, and the right panel shows the ratio of the two. The two outflow candidates are shown in red (NGC\,4424) and yellow (NGC\,4694), while to comparison galaxies of similar mass and inclination, IC\,1954 and NGC\,4298, are shown in green and blue, respectively.  All 18 galaxies in the PHANGS sample with inclinations greater than 55\textdegree\ are also shown as gray lines, with the median of this comparison sample shown as a black line.  \textit{Middle Row:} MIRI F770W$_{\rm SS}$, NIRCam F300M, and the ratio of the two maps of the primary comparison galaxy IC\,1954, on the same scales as Figures~\ref{fig:MW_ngc4424} and \ref{fig:MW_ngc4694}.  \textit{Bottom Row:} Same as previous, for comparison galaxy NGC\,4298.  For both comparison galaxies, the PAH emission (measured in F770W$_{\rm SS}$) and starlight (measured by F300M) basically trace the same structures.}
    \label{fig:Comparison}
\end{figure*}

To test whether these structures are signs of outflows or stripped gas, we compare our outflow candidates to other high inclination galaxies in the PHANGS-JWST sample. We select the 18 galaxies in the PHANGS-JWST sample with inclinations greater than 55\textdegree\ and a wide enough field of view to see above and below the stellar disk \footnote{NGC\,3239 was also excluded from this analysis due to its irregular structure, producing the final list of 18 galaxies.}.  Appendix~\ref{app:comp_sample} contains further details about all of the galaxies in the comparison sample, but to briefly summarize: this sample contains four additional Virgo Cluster galaxies, two galaxies with documented ram pressure stripping, and several galaxies currently undergoing tidal interactions with a nearby companion, making it a representative sample of other potential methods for removing or perturbing gas in a galaxy besides outflows.  Additionally, images of IC\,1954 and NGC\,4298 are shown in the lower panels of Figure~\ref{fig:Comparison} as representative examples of the full comparison sample due to their similar inclinations ($i\sim57\degree,~59\degree$) and field of view of the MIRI coverage reaching a similar height of the disk ($0.8\times R_{25}$), so we can be certain that any similar outflow structures would be included in the observations.  In Figure~\ref{fig:Comparison}, we show images of IC~1954 and NGC~4298 in F770W$_{\rm SS}$,  F300M, and the ratio of these two filters in the central (IC\,1945) and lower (NGC\,4298) panels.  As with all the galaxies in our comparison sample, there are no clear PAH-bright components extending off the disk, and the F770W$_{\rm SS}$ and F300M show similar structures. 

We plot the normalized profiles of F770W$_{\rm SS}$, F300M, and their ratio as a function of projected distance from the plane of the galaxy, $b$, for the two outflow hosts and all the galaxies in the comparison sample in the top panels of Figure~\ref{fig:Comparison}.  Each profile is normalized using the central value of the given intensity or ratio.  While the comparison sample and the outflow candidates show similar exponentially decreasing trends in F300M, the profiles of the F770W$_{\rm SS}$ data are significantly flatter in the two outflow candidates, indicating the presence of extended PAH emission off the plane of the galaxy. This difference in behavior is made clear in the top right panel of Figure~\ref{fig:Comparison}, which shows F770W$_{\rm SS}$/F300M as a function of distance from the plane.  While the comparison galaxies either remain fairly flat or show decreasing trends in this ratio (see gray lines for individual galaxies, or black line for the median of the full comparison sample), the two outflow candidates display an increasing ratio with distance from the plane.  This highlights the presence of PAH emission observed off the planes of these galaxies, tracing the outflowing gas.

 \section{Results}

\subsection{Comparison to Multi-wavelength Data}
\label{sec:mw_comp}

\subsubsection{JWST Photometry}
The outflows are primarily detected in the F770W band, tracing PAHs, but are also visible in the F2100W images of both galaxies.  This suggests that there are additional larger dust grains also being removed by these outflows.  Notably, in both galaxies, the areas classified as outflows have a slightly higher average F770W$_{\rm{SS}}$/F2100W than the stellar disks.  This could indicate that the PAH/dust ratio is slightly lower in the star-forming main disk than in the outflowing gas, but could also be due to differences in the radiation field heating the PAHs and dust in the disk compared to the outflowing gas, as a softer radiation field could still excite the PAHs without heating the dust as significantly, lowering the F2100W value \citep[see e.g. the models of][]{Draine2014}.

None of the NIRCam filters shows any sign of outflowing material in either galaxy.  Notably, this includes the F335M filter, which includes the 3.3~$\mu$m PAH emission feature typically associated with smaller PAH molecules \citep{Draine2021}, and the F187N filter, which contains the Paschen~$\alpha$ line, shown in the right-most bottom panels of Figures~\ref{fig:MW_ngc4424} (NGC\,4424) and \ref{fig:MW_ngc4694} (NGC\,4694).  Previous studies of the 3.3~$\mu$m/7.7~$\mu$m ratio in a set of star--forming galaxies showed very low ($\sim 0.05$) ratios across the full disks \citep[][H. Koziol, in prep]{Chastenet2023a}, indicating that the 3.3~$\mu$m feature may simply be below the detection threshold of these observations.  As the average surface brightness at F770W in the outflows was $\sim1$~MJy/sr, the expected F335M fluxes would be 0.05~MJy/sr, the $1\sigma$ limit for our observations \citep[see][Table 1]{Williams2024}.  While we cannot rule out the presence of 3.35~$\mu$m emission, we can state that it is not enhanced in the outflow, suggesting at minimum a PAH population that is not dominated by the smallest PAHs. This matches the simulations described in \citet{Richie2025}, which showed the smallest dust grains are less likely to survive in outflows.  Similarly, the lack of detection in the F187N filter could be due to sensitivity limits of this filter.  

\subsubsection{MUSE IFU Observations}
Both galaxies have partial coverage by the MUSE IFU.  We examine the emission from several optical emission lines including H$\alpha$, [\ion{O}{3}]~5007~\AA, and [\ion{N}{2}]~6584~\AA.  In both galaxies, these optical emission lines do show some extended emission off of the plane, which can be seen in the SFR maps in the lower middle panel of Figure~\ref{fig:MW_ngc4424} and \ref{fig:MW_ngc4694}.  \citet{Boselli2018} also identified an ionized gas outflow from NGC\,4424, but does not include NGC\,4694 as an outflow host. These ionized gas outflows seem to be adjacent to, but not overlapping with, the cool outflows detected in the F770W band. 

\subsubsection{ALMA CO observations}
No outflows were detected in either galaxy in the work of \citet{Stuber2021}, which classified outflows based on velocity profiles of two circular regions including the central 300~pc and 2000~pc from ALMA CO(2-1) mapping as well as $pv$ maps along the major and minor axes.  We compare the CO(2-1) intensity map to the MIRI F770W$_{\rm{SS}}$ data and find similar structures across the stellar disk for both galaxies, but do not detect CO throughout the full F770W-identified filaments.  CO(2-1) is detected by ALMA in the region defined as the outflow in both sources, but for consistency, we use only the F770W$_{\rm{SS}}$ derived gas masses in this work as $I_{\textrm{CO} (2-1)}$ falls below the sensitivity limits of these ALMA observations for the edges of the outflow structures.  

\subsubsection{VIVA HI 21 cm Maps}
Multiple papers have described the extended 21~cm emission detected around NGC\,4424 and NGC\,4694 \citep{Chung2007, Sorgho2017, Lee2022}.  These observations show an $\sim$18~kpc long tail extending from only the south-east side of NGC\,4424 and a $\sim30$~kpc tail from only the south-west side of NGC\,4694.  These tails are an order of magnitude larger than the outflows described here (0.73--2.3~kpc, Table~\ref{tab:galprops}), and the VIVA \ion{H}{1} maps have angular resolutions of 20\arcsec\ and are therefore not high enough angular resolution to determine if the filaments outlined in the black contours in Figure~\ref{fig:MW_ngc4424} and \ref{fig:MW_ngc4694} are detected at 21~cm.

\subsection{Estimating Mass Outflow Rates in the MIR}
\label{sec:out_rate}
Using the high sensitivity of the JWST-MIRI observations, we provide an order of magnitude estimate of the mass outflow rates in these two galaxies.  To do so, we first convert the F770W$_{\rm SS}$ map to a gas surface density map using the method described by \citet{Chown2025}.  We use the individual galaxy best-fit relationships to predict $I_{\rm{CO}(2-1)}$ in the CO bright part of the disks: 
$$\log_{10}I_{\textrm{CO}(2-1)} = m \times \log_{10}(\textrm{F770W}_{\textrm{ss}}-x_0)+b$$
with $m, b, x_0 = 0.816, 0.406, 0.480$ for NGC\,4424 and $0.821, 0.294, 0.360$ for NGC\,4694.  We use the predicted $I_{\rm{CO}(2-1)}$ instead of the ALMA maps to leverage the higher sensitivity of the MIRI observations, detecting even the low surface brightness edges of the outflow filaments that fall below the sensitivity of the ALMA observations, as discussed in Section~\ref{sec:mw_comp}.

With these F770W$_{\rm SS}$-scaled gas maps, the total mass of the outflows is estimated using the contours defined in Section~\ref{sec:id}.  We convert from predicted F770W$_{\rm SS}$-predicted $I_{\rm{CO}(2-1)}$ to a gas mass using a single $\alpha_{\rm{CO}}$ of $0.8$~M$_\odot$~pc$^{-2}~({\rm K\,km/s})^{-1}$, which is often used for starburst galaxies and outflows, and an $R_{21}$ of 0.65 \citep{Bolatto2013, Stuber2021}.  It should be noted that while the method of converting from MIRI photometry to a total gas mass has inherit scatter, the largest uncertainty in this calculation will come from the choice of $\alpha_{\rm{CO}}$, which could change the calculated gas mass by up to a factor of four.  Assuming a maximum and minimum $\alpha_{\rm{CO}}$ of the Milky Way $\alpha_{\rm{CO}}=4.35$~M$_\odot$~pc$^{-2}~({\rm K\,km/s})^{-1}$ and the optically thin limit of $\alpha_{\rm{CO}}= 0.35$~M$_\odot$~pc$^{-2}~({\rm K\,km/s})^{-1}$ respectively \citep[both from][]{Bolatto2013}, we get values for the gas mass of NGC\,4424: $\log_{10}M[$M$_\odot]=7.11^{7.85}_{6.76}$ and NGC\,4694: $\log_{10}M[$M$_\odot]=6.37^{7.10}_{6.01}$. 

The mass outflow rate is given by
$$\dot{M} = \frac{M \times v}{l},$$
where $v$ is the velocity of the outflow, $M$ is the gas mass of the outflow (as calculated above), and $l$ is the vertical extent of the outflow assuming a relatively constant density.  
Examining the moment 1 maps of the ALMA data described in Section~\ref{sec:data}, we find velocities of 13~km~s$^{-1}$ for NGC\,4424 and 39~km~s$^{-1}$ for NGC\,4694.  Assuming the maximal case that these velocities are exactly perpendicular to the plane of each galaxy yields a total velocity of 25~km~s$^{-1}$ for NGC\,4424 and 80~km~s$^{-1}$ for NGC\,4694.  It should be noted that less than half the pixels within the outflows have strong enough CO detections for a robust measurement of the velocity making these measurements highly uncertain.  If we instead assume a typical outflow velocity for galaxies of a similar star-formation rate from \citet{Stuber2021}, we would expect outflow velocities of $\sim$100-130~km~s$^{-1}$, although none of the galaxies within the sample of confirmed outflows have a star formation rate as low as NGC\,4694. Additional measurements of the ionized gas outflow in NGC\,4424 found it to have a velocity of 500~km~s$^{-1}$, much higher than either the velocity measured from the CO emission or the outflows measured in \citet{Stuber2021}.  This could be due to the $\sim3\times$ larger radius of the ionized gas outflow than the radii of the outflows we present here.  Based on these measurements, we assume a velocity of $v =80\pm20$~km~s$^{-1}$ to span the measured and deprojected velocity of NGC\,4694 and the velocities of the lowest SFR outflows presented in \citet{Stuber2021}.

$l$ is measured using the largest radial extent of the red contours shown in Figures~\ref{fig:MW_ngc4424} and \ref{fig:MW_ngc4694}, and is found to be between 1 and 2~kpc in both sources.  Properties of the outflows are listed in Table~\ref{tab:galprops}.  As we do not know the exact geometry of the outflows, we do not correct for inclination, so $l$ is the length of the outflows as projected on the sky.  Using this method, we find the mass outflow rate of NGC\,4424 is $\sim 0.4^{+2.0}_{-0.3}$~M$_{\odot}$\,yr$^{-1}$ and NGC\,4694 is $\sim0.2^{+0.8}_{-0.15}$~M$_{\odot}$\,yr$^{-1}$. These mass outflow rates are close to what would be expected for a mass loading factor ($\dot{M}$/SFR) of $\sim$unity, matching the average value found in \citet{Stuber2021}, though these galaxies have both SFR and $\dot{M}$ about one order of magnitude lower than the typical value of 3~M$_{\odot}$~yr$^{-1}$ for the identified outflow hosts in that paper.

\section{Conclusions}
\label{sec:conc}
In this Letter, we have shown how MIRI observations can be used to characterize cool outflowing gas in two galaxies: NGC\,4424 and NGC\,4694.  These outflow candidates were measured by comparing the relatively smooth stellar disk to the clearly extended PAH and dust disk.  Both of these galaxies are small ($\log_{10}M_\star/M_\odot=9.9$) and highly inclined ($i>55$\textdegree), making our observing angle ideal for seeing the outflowing dust and gas, and have some signs of additional ionized gas outflows detected in optical emission lines \citep{Boselli2018}.  The outflowing gas in both sources has slightly elevated F770W$_{\rm{SS}}$/F2100W compared to their disks, either signaling increased PAH fraction or changes to the radiation field heating the dust.  While cool gas outflows have been detected in low mass galaxies before \citep[e.g. SMC, NGC~1569, NGC~55, NGC~5253;][]{McCormick2013}, the low surface brightness of the outflowing gas has made it difficult to observe the lower mass, cool outflows in smaller galaxies at larger distances.  The new infrared observations from JWST provide the link between the dust and gas in these outflows and the ability to resolve the structures present in the outflowing gas.

This work shows that mid-infrared observations with JWST hold the potential for discovery of galaxy-scale outflows in highly inclined sources where gas kinematics are poorly suited for outflow detection.  This result demonstrates the power of using PAH-dominated MIR imaging with JWST to trace gas in galaxies \citep{Leroy2023, Chown2025}, opening a new window for studying low mass outflows that would be undetected even with ALMA. This work signals the potential discovery space for finding dusty outflows in lower mass galaxies. 

\begin{acknowledgments}
The authors would like to thank the anonymous referee for their careful consideration of this paper and the detailed feedback that improved the analysis.  HAP acknowledges support from the National Science and Technology Council of Taiwan under grant 113-2112-M-032-014-MY3.
JC acknowledges funding from the Belgian Science Policy Office (BELSPO) through the PRODEX project ``JWST/MIRI Science exploitation'' (C4000142239).
MB acknowledges support from the ANID BASAL project FB210003. This work was supported by the French government through the France 2030 investment plan managed by the National Research Agency (ANR), as part of the Initiative of Excellence of Université Côte d’Azur under reference number ANR-15-IDEX-01. OE acknowledges funding from the Deutsche Forschungsgemeinschaft (DFG, German Research Foundation) -- project-ID 541068876.

This work has been carried out as part of the PHANGS collaboration. This work is based on observations made with the NASA/ESA/CSA {\it JWST}. The data were obtained from the Mikulski Archive for Space Telescopes at the Space Telescope Science Institute, which is operated by the Association of Universities for Research in Astronomy, Inc., under NASA contract NAS 5-03127 for {\it JWST}. These observations are associated with programs 2107 and 3707. The specific observations analyzed can be accessed via \dataset[DOI: 10.17909/ew88-jt15]{https://doi.org/10.17909/ew88-jt15}.

This work is also based on observations collected at the European Southern Observatory under ESO programs: 097.D-0408 (PI J. Anderson) and 110.244E (PI L. Cortese)

This paper makes use of the following ALMA data:
ADS/JAO.ALMA\#2017.1.00886.L.
ALMA is a partnership of ESO (representing its member states),
NSF (USA), and NINS (Japan), together with NRC (Canada),NSC and ASIAA (Taiwan), and KASI (Republic of Korea), in cooperation with the Republic of Chile. The Joint ALMA Observatory is operated by ESO, AUI/NRAO, and NAOJ. The National Radio Astronomy Observatory is a facility of the National Science Foundation operated under cooperative agreement by Associated Universities, Inc.

\end{acknowledgments}

%

\vspace{5mm}
\facilities{HST(STIS), Swift(XRT and UVOT), AAVSO, CTIO:1.3m,
CTIO:1.5m,CXO}


\software{astropy \citep{2013A&A...558A..33A,2018AJ....156..123A},
          }



\appendix
\section{Comparison Sample Properties}
\label{app:comp_sample}

In this Appendix, we provide a brief summary of the properties of the galaxies in the comparison sample.  The galaxies in this sample are all part of the PHANGS-JWST Cycle 2 Treasury Program, and were selected primarily due to their high inclinations ($i<55$\textdegree).  The galaxies in this sample are used to exemplify how ram pressure stripping or recent gravitational interactions alone cannot produce the elevated off-plane PAH/starlight ratios observed in our two outflow candidate galaxies.  It is therefore essential we consider the environment and recent history of each galaxy in the comparison sample.  Importantly, nearly all of the galaxies in this sample are members of larger galaxy groups, as specified in the Group Member column of Table~\ref{tab:app}, with four Virgo Cluster galaxies.  Two of the Virgo Cluster members also have documented ram pressure stripping: NGC\,4569 \citep{Boselli2016} and NGC\,4654 \citep{Sofue2003}. Two of the comparison sample are currently undergoing tidal interactions with a nearby companion (NGC\,4907 and NGC\,4298), and NGC\,4826 recently merged with a SMC-like companion \citep{Smercina2023}.  One has documented inflowing gas \citep[NGC\,4536][]{daSilva2024} and another has an extended UV disk \citep[NGC\,2090][]{Thilker2007}.  Despite this wide range of environmental influences, none of our comparison sample show the same elevated PAH to stellar emission off the plane of the disk as our two outflow host candidate galaxies.

\begin{deluxetable}{lcccccc} 
    \caption{Properties of Comparison Sample Galaxies} 
    \tablehead{Galaxy & $i$[\textdegree] & $D$ [Mpc] & log$_{10}$M$_{\star}$ & log$_{10}$SFR & Group& Tidal\\ 
    & & & & & Membership & Tail? }
    \startdata 
    \textbf{IC1954} &  57.1 & 12.8 & 9.7 & $-$0.44 & LGG093 & ...\\ 
    NGC1546 & 70.3 & 17.7 & 10.4 & $-$0.08 & Dorado & ... \\ 
    NGC1559 & 65.4 & 19.4 & 10.4 & 0.58 & Dorado & ... \\ 
    NGC1792 & 65.1 & 16.2 & 10.6 & 0.57 & NGC1808 Group & ... \\ 
    NGC1809 & 57.6 & 20.0 & 9.8 & 0.76 & ... & ... \\ 
    NGC2090 & 64.5 & 11.7 & 10.0 & $-$0.39 & ... & ... \\ 
    NGC2903 & 66.8 & 10.0 & 10.6 & 0.49 & NGC2903 Group & ... \\ 
    NGC3137 & 70.3 & 16.4 & 9.9 & $-$0.31 & LGG189 & ... \\ 
    NGC3511 & 75.1 & 13.9 & 10.0 & $-$0.09 & ... & ... \\ 
    NGC3521 & 68.8 & 13.2 & 11.0 & 0.57 & ... & ... \\ 
    \textbf{NGC4298} & 59.2 & 14.9 & 10.0 & $-$0.34 & Virgo & ... \\ 
    NGC4536 & 66.0 & 16.3 & 10.4 & 0.54 & Virgo & ... \\ 
    NGC4569 & 70.0 & 15.8 & 10.8 & 0.12 & Virgo & \checkmark \\ 
    NGC4654 & 55.6 & 22.0 & 10.6 & 0.58 & Virgo & \checkmark \\ 
    NGC4731 & 64.0 & 13.3 & 9.5 & $-$0.22 & LGG314 & ... \\ 
    NGC4781 & 59.0 & 11.3 & 9.6 & $-$0.32 & LGG307 & ... \\ 
    NGC4826 & 59.1 & 4.4 & 10.2 & $-$0.70 & ... & ... \\ 
    NGC4951 & 70.2 & 15.0 & 9.8 & $-$0.45 & LGG314 & ... \\ 
    \enddata
\tablecomments{Properties of the 18 comparison galaxies, selected from the PHANGS-JWST data.  Comparison galaxies are selected based on their inclinations  ($i>55$\textdegree) and the extent of the JWST coverage above and below the stellar plane.  Inclinations ($i$) and distances ($D$) are from \citet{Anand2021} and M$_{\star}$ and SFR are from \citet{Leroy2019}. Galaxy cluster membership from the \cite{Tully2015} catalog.  Galaxies with identified tidal tail from \citet{Lee2022} are marked with a $\checkmark$. Primary comparison galaxies are listed in bold.}
\label{tab:app}
\end{deluxetable}

\bibliography{main}{}
\bibliographystyle{aasjournal}



\end{document}